# Two-photon absorption of time-frequency-entangled photon pairs by molecules: the roles of photon-number correlations and spectral correlations

*Michael G. Raymer,[1,2]\* Tiemo Landes,[1,2] Markus Allgaier,[1,2] Sofiane Merkouche,[1,2] Brian J. Smith,[1,2] Andrew H. Marcus [2,3]*

[1.] Department of Physics, University of Oregon, Eugene, OR 97403, USA

[2.] Oregon Center for Optical, Molecular and Quantum Science, University of Oregon, Eugene, OR 97403, USA

[3.] Dept of Chemistry and Biochemistry, University of Oregon, Eugene, OR 97403, USA

\* Corresponding author: raymer@uoregon.edu

Keywords: quantum light spectroscopy, two-photon absorption, two-photon absorption, quantum enhancement, spontaneous parametric down-conversion

## Abstract:

While two-photon absorption (TPA) and other forms of nonlinear interactions of molecules with isolated time-frequency-entangled photon pairs (EPP) have been predicted to display a variety of fascinating effects, their potential use in practical quantum-enhanced molecular spectroscopy requires close examination. This paper presents a detailed theoretical study of quantum-enhanced TPA by both photon-number correlations and spectral correlations, including an account of the deleterious effects of dispersion. While such correlations in EPP created by spontaneous parametric down conversion can increase the TPA rate significantly in the regime of extremely low optical flux, we find that for typical molecules in solution this regime corresponds to such low TPA event rates as to be unobservable in practice. Our results support the usefulness of EPP spectroscopy in atomic or other narrow-linewidth systems, while questioning the efficacy of such approaches for broadband systems including molecules in solution.

## 1. Introduction

In the past decade many theoretical studies have proposed quantum advantages in spectroscopy by the use of time-frequency-entangled photon pairs (EPP). For review see [1,2,3,4] These include, for example, proposals for virtual-state spectroscopy [5], Raman spectroscopy [6], and multi-dimensional optical spectroscopy [7,8]. A useful benchmark test of the current thinking on spectroscopy using time-frequency EPP is provided by two-photon absorption (TPA) in the absence of resonant intermediate states. This paper examines in detail a first-principles theory of TPA and offers the conclusion that the quantum advantage of EPP for spectroscopy is likely not achievable for typical molecules in solution. Nevertheless, our results support the utility of EPP spectroscopy in atomic or other narrow-linewidth systems.

The basis of the current thinking on EPP spectroscopy goes back to the seminal theoretical papers by Gea-Banacloche [9] and by Javanainen and Gould. [10] These initial studies showed that in the regime of 'isolated' EPP—defined as the regime of extremely low flux where not more than two photons on average impinge on the molecule within the field's coherence time and not more than two photons impinge on the molecule within its response time — the TPA probability for a single molecule scales linearly with the photon flux (since the photons come in pairs). Moreover, for EPP created by spontaneous parametric down conversion (SPDC), an increased bandwidth of the EPP field does not decrease the absorption probability. This effect occurs because the frequencies of the two EPP photons are not random, but rather anticorrelated so that they sum to the constant narrow-band frequency of the pump laser. We refer to this effect as enhancement due to spectral correlation. The first of these predictions was verified in an experiment using narrow-band EPP to excite a two-photon transition in a vapor of atomic cesium. [11] The second prediction found experimental support in atomic systems for a slightly different scenario: TPA in the high flux regime where the linear scaling of the TPA rate is lost, but the enhancement is retained by frequency anticorrelation. [12] Subsequent studies using molecular solutions reported orders-of-magnitude enhancement of TPA using EPP [13, 14], but recently those studies have been called into question. [15, 16]

To see the scope of the challenge, consider a 1-cm thick solution of typical dye molecules (e.g., Rhodamine 6G) with concentration 2 millimolar (i.e., $1.2 \times 10^{18} cm^{-3}$). The conventional TPA cross section at wavelength 1064 nm is of order $9 \times 10^{-50} cm^4 s / photon^2$. Thus, illuminating with a monochromatic 1064-nm continuous-wave laser beam with transverse cross-sectional area $10^{-6}$ cm$^2$, and power 20 nW ($10^{11}$ $photons/s$), yields a conventional TPA rate of about $1 \times 10^{-3}$ $events/s$, and assuming a collection-plus-detection efficiency of 0.01, predicts a TPA fluorescence count rate of about $1 \times 10^{-5} /s$. For this estimate the mean photon flux was chosen to correspond to be near the maximum flux for an EPP beam with bandwidth $1 \times 10^{13}$ $Hz$ to remain in the isolated-pair regime. Without many orders of magnitude enhancement by the special properties of entangled light, such a fluorescence signal would be unobservable. Attempting to observe TPA in transmission would involve measurement of one part in $10^{17}$ attenuation of a transmitted beam and would be similarly difficult to achieve. Furthermore, the use of sub-ps pulses of EPP would not increase the excitation probability as long as the EPP pulse remains in the isolated-pair regime.

The questions addressed in this paper are: 1. How large is the predicted enhancement of TPA by time-frequency entanglement of isolated EPP? and 2. Is it practical experimentally to achieve the predicted enhancements in molecular TPA? To answer these questions, we provide a unified treatment of TPA under a wide range of absorption conditions for both coherent laser light, and isolated EPP. We consider both pulsed and continuous-wave (CW) sources, and we calculate explicitly the degree of enhancement that can be achieved using EPP relative to coherent-state light.

Based on this careful analysis, we conclude that while TPA using isolated time-frequency-entangled EPP can, in principle, be enhanced by several orders of magnitude relative to coherent-state absorption, the regime where this effect is significant in typical molecules corresponds to such low TPA event rates as to be unobservable in practice. This conclusion is corroborated by our laboratory experiments, which we report elsewhere [15], and by a separate, independent experimental study. [16]

The above conclusion can be summarized as follows: There is no single, unique definition of a quantum enhancement factor for TPA—the amount of enhancement for quantum light depends on which 'classical' state of light being compared to: long-pulsed or short-pulsed. For the cases of CW or long-pulsed illumination, we can define a quantum enhancement factor QEF as the ratio of broad-band EPP absorption probability to narrow-band coherent-state absorption probability for *equal* average fluxes of EPP and coherent-state photons. For this case we obtain an especially simple form for the enhancement factor: $QEF = B\chi / F_{EPP}$, where $B$ is the bandwidth of the EPP, $\chi$ is a factor of order one depending on the line shape of the EPP spectrum, and $F_{EPP}$ is the flux (photons per second) in the EPP stream. The factor $B / F_{EPP}$ is a measure of the number of temporal (time-frequency) modes that impinge on the molecule in a time $1/F_{EPP}$, and can be very large in the limit that $F_{EPP}$ is very small. In this regime of large enhancement there is only one pair of entangled photons spread over the time $1/F_{EPP}$, and these photons are tightly correlated in time (to within roughly $1/B$). We show that given the magnitude of TPA cross sections for typical molecules, the large enhancement occurs only in the regime of vanishing signal count rate. In the example given above, the QEF would have the value $QEF \sim B / F_{EPP} = 10^{13}\ Hz / 10^{11}\ Hz = 100$, very far from being large enough to enable detection of signal.

We find that for the case of ultra-short excitation pulses, if the EPP bandwidth is comparable to the inverse duration of the pulse, there is no significant quantum enhancement arising from spectral correlation. In this case, the EPP state does not contain significant amounts of time-frequency entanglement. Furthermore, for a narrow transition line and broad EPP spectrum we find that the excitation probability depends very much on the amount and nature of entanglement in the exciting light. Nevertheless, in any case the separate effect of photon number correlation persists since the EPP always arrive in pairs.

Dispersion of the EPP field by passage through typical linear-optical elements is known to decrease the efficacy of TPA by EPP. Our formalism allows incorporating rigorously the effects of dispersion in TPA, leading to a simple formula quantifying the expected decrease of the EPP-driven TPA probability caused by the increase of the two-photon correlation time.

Several theoretical studies with goals similar to ours have appeared previously but considered only a subset of the issues we treat. Dayan emphasized cases with no resonant intermediate states, which is also our focus, and also treated the case of overlapping EPP at high flux. [17] Schlawin and Buchleitner emphasized cases in which resonant intermediate states play a strong role in TPA. [18]

We note that when quoting TPA cross sections, we generally refer to the *conventional cross section* as measured by narrow-band coherent-state light. While Fei et. al. introduced an "entangled two-photon cross section," [19] we prefer to separate the properties of the molecular cross section from the properties of the light. We do, however, give the relation between our enhanced cross section and the entangled two-photon cross section of Fei et. al. (See Eq.(68).)

Throughout the remainder of the paper, we present derivations in compact form, and give additional details to these derivations as well as further examples in Appendices.

## 2. Theory of one- and two-photon absorption—General formalism

The theory of molecular interactions with quantized light, including TPA, has been treated numerous times; the traditional approach uses perturbation theory and posits a density of molecular or photonic states to arrive at the Fermi Golden Rule for the *conventional* cross section for TPA. [20, 21, 22] When dealing with coherent light pulses or light having time-frequency correlations, a more detailed treatment is needed, and several such treatments have appeared. [18,17, 1, 9, 10] It seems that while such treatments predict intriguing aspects of TPA using EPP, few if any considered carefully under what experimental conditions such effects could be observed.

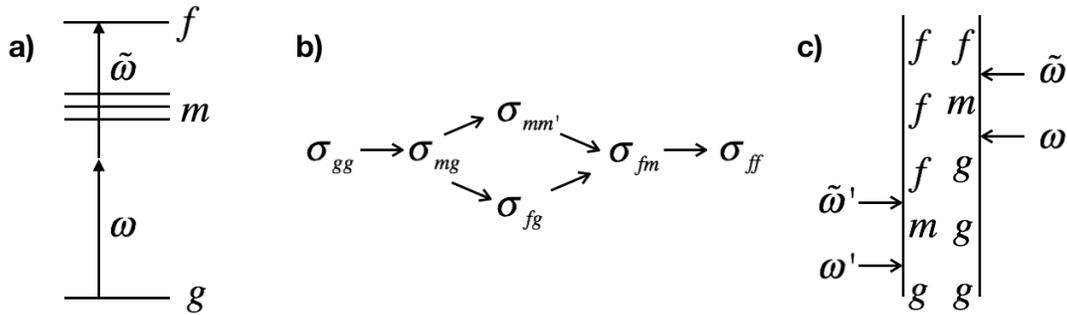

**Fig. 1** a) A single molecule with ground, intermediate, and final states labeled g, m, and f, with energies $E_i = \hbar\omega_i$. b) The perturbative sequence in which molecular transition operators $\hat{\sigma}_{ij} = |i\rangle\langle j|$ are excited. c) Double-sided Feynman diagram representing the dominant pathway when there are no resonant intermediate states.

In the literature, it is common to use a molecular density-matrix formalism to model absorption or nonlinear-optical response of multilevel systems. [23] While quantum states of light as the driver of absorption can be described by the density-matrix formalism [1], here we use a different formalism common in the quantum optics literature – the operator version of the optical Bloch equations. [24,25] These equations respect the quantum nature of the field via commutation relations involving raising and lowering operators for the field and the molecular states.

Consider a single molecule undergoing TPA, as in **Fig. 1** showing a set of molecular energy eigenstates satisfying $\hat{H}_M|i\rangle = \hbar\omega_i|i\rangle$. The optical Bloch equations track the time evolution of molecular-state transition-projection operators $\hat{\sigma}_{ij}(t) = |i\rangle\langle j|$, whose expectation values equal populations (that is probabilities) and 'coherences' via the familiar density-matrix elements

$\langle \hat{\sigma}_{ij} \rangle = \rho_{ji}$. The Heisenberg-picture equations of motion are $\partial_t \hat{\sigma}_{ij} = (1/i\hbar)[\hat{\sigma}_{ij}, \hat{H}]$, where the Hamiltonian is $\hat{H} = \hat{H}_M + \hat{H}_F - \vec{d} \cdot \vec{\varepsilon} \hat{E}$, where $H_F$ is the energy of the field, $\vec{d}$ is the electric-dipole operator, and $\vec{\varepsilon} E$ is the field operator with $\vec{\varepsilon}$ being its polarization vector. The equations of motion (without dephasing interactions) read explicitly:

$$\frac{d}{dt}\hat{\sigma}_{ij} = i\omega_{ij}\hat{\sigma}_{ij} + i\sum_m \left(\mu_{jm}\hat{\sigma}_{im} - \mu_{mi}\hat{\sigma}_{mj}\right)\hat{E} \tag{1}$$

with $\omega_{ij} = \omega_i - \omega_j$ and the electric dipole matrix elements are $\mu_{ij} = \vec{d}_{ij} \cdot \vec{\varepsilon}/\hbar$. In order to identify clearly the various quantum pathways leading to the populations of interest, we carry out explicitly the perturbative solution of the equations of motion, as indicated in **Fig. 1.** Details are given in **Appendix A**. We assume throughout that the initial state is $|g\rangle|\Psi\rangle$ for a pure state $|\Psi\rangle$ of the field, or for a mixed field state a density operator: $\hat{\rho} = |g\rangle\langle g| \otimes \hat{\rho}_{Field}$. The first-order probability to find the molecule in state $m$ (excited by single-photon absorption) is $P_m \equiv \rho_{mm}^{(2)}(\infty) = p_m + cc$, where the 'complex probability' is:

$$p_m = \mu_{mg}\mu_{gm} \int_{-\infty}^{\infty} dt_2 \int_{-\infty}^{t_2} dt_1 e^{-(\gamma_{mg} - i\omega_{mg})(t_2 - t_1)} \langle \hat{E}(t_1)\hat{E}(t_2) \rangle \tag{2}$$

in which molecular dephasing rates (from elastic collisions, not population damping) $\gamma_{mg}$ have been included using the prescription of Kubo. [23] The field correlation function is given by the expectation value:

$$\begin{aligned} C(t_1, t_2) &\equiv \langle \hat{E}(t_1)\hat{E}(t_2) \rangle \\ &= Tr_{Field}\left(\hat{\rho}_{Field}\hat{E}(t_1)\hat{E}(t_2)\right) \end{aligned} \tag{3}$$

Changing variables to $\tau = t_2 - t_1$ gives:

$$p_m = \mu_{mg}\mu_{gm} \int_{-\infty}^{\infty} dt_2 \int_0^{\infty} d\tau e^{-(\gamma_{mg} - i\omega_{mg})\tau} C(t_2 - \tau, t_2) \tag{4}$$

To evaluate the correlation function, we write the field operator as a sum of positive- and negative-frequency parts, $\hat{E}(t) = \hat{E}^{(+)}(t) + \hat{E}^{(-)}(t)$, where [26]

$$\hat{E}^{(+)}(t) = \int d\omega L(\omega)\hat{a}(\omega)e^{-i\omega t} \tag{5}$$

with $\hat{E}^{(-)} = \hat{E}^{(+)\dagger}$ and $L(\omega) = \sqrt{\hbar\omega/2\varepsilon_0 ncA_0}$. The frequency $\omega$ has units rad/s, $n$ is the refractive index of the medium, and $A_0$ is the effective beam area at the molecule's location. $\varepsilon_0$ is the vacuum permittivity, and $c$ is the speed of light. The commutator for creation and annihilation operators is $[\hat{a}(\omega'), \hat{a}^\dagger(\omega)] = 2\pi\delta(\omega'-\omega)$. And we use the shorthand notation $\dbar\omega = d\omega/2\pi$.

Because Eq.(5) represents a single spatial mode, it does not describe spatial correlation or entanglement. The photons are assumed distributed across the beam independently, as is the case if their generation takes place in a single-mode wave guide supporting a single polarization or is passed through a single-mode optical fiber and linear polarizer. Therefore, in the case of EPP generated by SPDC, the 'entanglement area' is the transverse area of the EPP beam at the molecule's location, unlike in treatments where transverse spatial correlations are considered. [19] For a wide-area beam, such correlations can localize photon pairs to transverse areas much smaller than the overall beam area but cannot localize pairs to an area much smaller than the diffraction-limited focus of a well-designed optical system.

For near-resonant one-photon-absorption, the rotating-wave approximation is valid, meaning we approximate:

$$C(t_1, t_2) \simeq \left\langle \hat{E}^{(-)}(t_1)\hat{E}^{(+)}(t_2) \right\rangle \tag{6}$$

Using Eqs. (5) and (6), the integral in Eq.(4) can be carried out, as in **Appendix A**, to yield:

$$p_m \simeq |\mu_{gm}|^2 L_0^2 \int \dbar\omega \frac{\left\langle \hat{a}^\dagger(\omega)\hat{a}(\omega)\right\rangle}{\gamma_{mg} - i(\omega_{mg} - \omega)} \tag{7}$$

where $L_0 = \sqrt{\hbar\omega_0/2\varepsilon_0 ncA_0}$ and $\omega_0$ is the (central) carrier frequency of the exciting light. This result is, of course, just the overlap of the exciting field spectrum with the absorption profile and is independent of any spectral phase in the field. Examples of one-photon-absorption by specific states of the field are given in **Appendix B**.

Next consider two-photon absorption with non-resonant intermediate states, leading to population in the final state $|f\rangle$, therefore the dominant pathway is $\sigma_{gg} \to \sigma_{mg} \to \sigma_{fg} \to \sigma_{fm(fm')} \to \sigma_{ff}$. Thus, we neglect the alternate pathway shown in **Fig. 1** that passes through 'real' intermediate-state populations $\sigma_{mm}$ or coherences $\sigma_{mm'}$. This approximation is clearly valid if all intermediate states lie at energies above that of state $f$, and may be valid even if they lie below state $f$ but are far from being one-photon resonant.

The population $\rho_{ff}(\infty) = \langle \sigma_{ff}(\infty) \rangle$ in state $|f\rangle$ following the exciting pulse (which might be a time window taken arbitrarily from a continuous-wave exciting beam) is given by the fourth-order expression:

$$P_f \equiv \rho_{ff}^{(4)}(\infty) = p_f + cc \tag{8}$$

where the 'complex probability' is:

$$p_f = \sum_{m,m'} \mu_{fm}\mu_{mg}\mu_{m'f}\mu_{gm'} p_{mm'} \tag{9}$$

with

$$p_{mm'} = \int_{-\infty}^{\infty} dt \int_0^{\infty} dr \int_0^{\infty} ds \int_0^{\infty} d\tau\, e^{-(\gamma_{fm}-i\omega_{fm})r} e^{-(\gamma_{fg}-i\omega_{fg})s} e^{-(\gamma_{m'g}-i\omega_{m'g})\tau} C^{(4)} \tag{10}$$

in which

$$C^{(4)} = \left\langle \hat{E}(t-r-s-\tau)\hat{E}(t-r-s)\hat{E}(t-r)\hat{E}(t) \right\rangle \tag{11}$$

This result is consistent with those in [9, 10, 2].

Again, it is instructive to write the result in the spectral domain. Applying the rotating-wave approximation, valid near or at two-photon resonance (see **Appendix A**), and considering that the exciting field is far from resonance with any intermediate states ($\tilde{\omega} \neq \omega_{fm}, \omega' \neq \omega_{m'g}$), we obtain:

$$p_{mm'} \simeq \frac{L_0^4}{(-\omega_{fm}+\omega_0)(\omega_{m'g}-\omega_0)} \int d\omega \int d\tilde{\omega} \int d\omega' \frac{\left\langle \hat{a}^\dagger(\omega')\hat{a}^\dagger(\omega+\tilde{\omega}-\omega')\hat{a}(\omega)\hat{a}(\tilde{\omega}) \right\rangle}{\gamma_{fg} - i\omega_{fg} + i\omega + i\tilde{\omega}} \tag{12}$$

where $\omega_0$ is the central frequency of the EPP spectrum. Here and throughout the paper we define a real quantity that is proportional to the conventional two-photon-absorption cross section (recall $\mu_{ij} = \vec{d}_{ij} \cdot \vec{\varepsilon}/\hbar$):

$$\Sigma^{(2)} = \sum_{m,m'} \frac{\mu_{fm}\mu_{mg}\mu_{m'f}\mu_{gm'}}{(-\omega_{fm}+\omega_0)(\omega_{m'g}-\omega_0)} \tag{13}$$

Then we can write Eqs.(9) with (12) as:

$$p_f = \Sigma^{(2)} L_0^4 \int d\omega \int d\tilde{\omega} \int d\omega' \frac{\left\langle \hat{a}^\dagger(\omega')\hat{a}^\dagger(\omega+\tilde{\omega}-\omega')\hat{a}(\omega)\hat{a}(\tilde{\omega}) \right\rangle}{\gamma_{fg} - i\omega_{fg} + i\omega + i\tilde{\omega}} \tag{14}$$

where the dimensionless factor $\Sigma^{(2)} L_0^4$ is proportional to the conventional two-photon absorption (excitation) cross section, as discussed in **Appendix G**. Note that the transverse area $A_0$ of the incident beam at the molecule's location is accounted for in the factor $L_0^4$, which is common to all expressions.

## 3. Classical and quantum states of light

Here we summarize the basic properties of coherent states, single-photon states, and two-photon states. For convenience, we consider the incident light to be in pulses of finite duration. In the case of CW excitation, we imagine the field to be made of a series of rectangular pulses with constant mean power and duration $T$, as in **Fig.2(a)**. When comparing to TPA with short pulses of SPDC or coherent-state light, it suffices to consider only a single pulse occupying the same interaction time window $T$, as in **Figs.2(b) and (c)**.

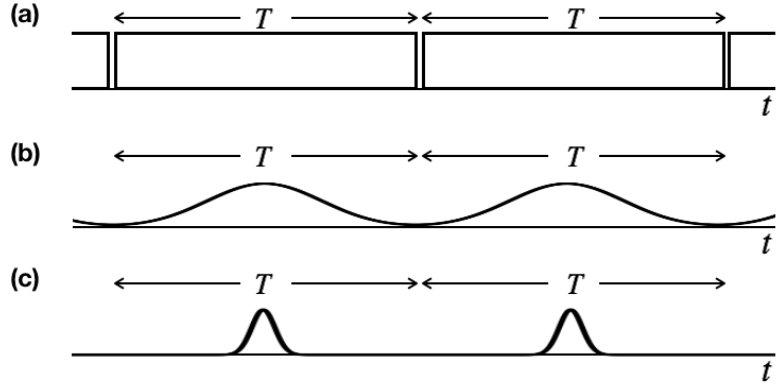

Fig. 2 Rectangular and Gaussian pulse trains with period $T$.

### 3.1. Coherent state

A pulsed coherent state $|\alpha\rangle$ with spectral amplitude $\alpha(\omega)$ satisfies $\hat{a}(\omega)|\alpha\rangle = \alpha(\omega)|\alpha\rangle$ at each frequency, or equivalently in the time domain:

$$\hat{E}^{(+)}(t)|\alpha\rangle = L_0 A(t) e^{-i\omega_0 t}|\alpha\rangle \tag{15}$$

where:

$$A(t) = \int d\omega\, \alpha(\omega) e^{-i(\omega-\omega_0)t} \tag{16}$$

$A(t)$ is a slowly varying envelope and $\omega_0$ is the (central) carrier frequency. The mean number of photons in the pulse is the time-integral of the flux $|A(t)|^2$:

$$N = \int |\alpha(\omega)|^2 d\omega = \int |A(t)|^2 dt \tag{17}$$

It is useful to define a normalized spectral amplitude $\phi(\omega)$, called a temporal mode, [27] by $\alpha(\omega) = \alpha_0 \phi(\omega)$, where $\int |\phi(\omega)|^2 d\omega = 1$. Then $N = |\alpha_0|^2$.

### 3.2. Single-photon state

A single-photon state of a particular temporal mode (coherent wave packet) $\tilde{\varphi}(\omega)$ is described by: [27]

$$|\varphi\rangle = \int d\omega \, \tilde{\varphi}^*(\omega) \hat{a}^\dagger(\omega) |vac\rangle \tag{18}$$

where the spectral density is normalized as $\int d\omega \, |\tilde{\varphi}(\omega)|^2 = 1$. Such a state can be created by, for example, heralded SPDC. [28]

### 3.3. Entangled photon pair state

Collinear Type-0 or Type-I spontaneous parametric down conversion pumped by a pulse of finite duration can be designed to occur into a single spatial-and-polarization mode [29]; then the state is described by:

$$|\Psi\rangle = \sqrt{1-\varepsilon^2} |vac\rangle + \varepsilon \int d\omega \int d\tilde{\omega} \, \psi(\omega,\tilde{\omega}) \hat{a}^\dagger(\omega) \hat{a}^\dagger(\tilde{\omega}) |vac\rangle + ... \tag{19}$$

The probability that a given pulse contains a photon pair is $\varepsilon^2 \ll 1$. We neglect higher-order terms representing generation of multiple pairs in order to satisfy our assumption of isolated EPP interacting with the molecule. The joint-spectral amplitude (JSA) $\psi(\omega,\tilde{\omega})$ is determined by the spectrum of the pumping laser pulse and the phase-matching properties of the nonlinear crystal used as second-order nonlinear medium. [30] It is normalized: $\int d\omega \int d\tilde{\omega} \, |\psi(\omega,\tilde{\omega})|^2 = 1$. The considerations discussed around Fig. 2 reveal the mean photon flux (twice the pairs flux), time-averaged over the whole interval $T$, to be $F_{EPP} = 2\varepsilon^2/T$, for either CW or pulsed cases.

For a CW or quasi-CW pump with central frequency $\omega_P$, the JSA is largest for $\omega + \tilde{\omega} = \omega_P$, that is, the frequencies are anti-correlated. An important property of the JSA in cases for which there are no distinguishing labels on the photon creation operators in Eq.(19), other than frequency, is the symmetry $\psi(\omega,\tilde{\omega}) = \psi(\tilde{\omega},\omega)$. Type-II SPDC, where the JSA need not be symmetric, is discussed in **Section 5**.

## 4. Two-photon absorption (TPA) with coherent states

Here we give expressions for two-photon molecular excitation by a coherent state, without resonant intermediate states. Another 'classical' example is TPA by a short coherent pulse in the impulsive limit, which is discussed in further detail in **Appendix C**.

We start from Eq.(14), where the frequency correlation function is denoted:

$$\tilde{C}_\omega = \left\langle \hat{a}^\dagger(\omega')\hat{a}^\dagger(\omega+\tilde{\omega}-\omega')\hat{a}(\omega)\hat{a}(\tilde{\omega}) \right\rangle \tag{20}$$

Consider a pulsed coherent state satisfying $\hat{a}(\omega)|\alpha\rangle = \alpha(\omega)|\alpha\rangle$ across the pulse spectrum. Then the correlation function is the product:

$$\tilde{C}_\omega = \alpha^*(\omega')\alpha^*(\omega+\tilde{\omega}-\omega')\alpha(\omega)\alpha(\tilde{\omega}) \tag{21}$$

which yields the complex probability:

$$p_f = \Sigma^{(2)} L_0^4 \int d\omega \int d\tilde{\omega} \int d\omega' \frac{\alpha^*(\omega')\alpha^*(\omega+\tilde{\omega}-\omega')\alpha(\omega)\alpha(\tilde{\omega})}{\gamma_{fg} - i\omega_{fg} + i(\omega+\tilde{\omega})} \tag{22}$$

Then, using $\alpha(\omega) = \alpha_0 \phi(\omega)$ and changing variables to $x = (\omega + \tilde{\omega} - 2\omega_0); z = \omega - \omega_0; z' = \omega' - \omega_0$, we obtain:

$$p_f = |\alpha_0|^4 \Sigma^{(2)} L_0^4 \int dx \frac{|K_{coh}(x)|^2}{\gamma_{fg} - i\omega_{fg} + i(2\omega_0 + x)} \tag{23}$$

where:

$$K_{coh}(x) = \int dz\, \phi(\omega_0 + z)\phi(\omega_0 + x - z) \tag{24}$$

which is a convolution representing the different combinations of frequencies that effectively create excitation near $\omega + \tilde{\omega} = 2\omega_0$. We see the expected quadratic scaling with $N$, via $|\alpha_0|^4 = N^2$. The qualitative understanding of Eq.(23) is described below.

### 4.1 Coherent state TPA, Gaussian pulse

As an example of Eq.(23), consider a Gaussian pulsed coherent-state having a spectrum with bandwidth $\sim \sigma$, i.e., $\alpha(\omega) = \alpha_0 (\sigma^2/2\pi)^{-1/4} \exp(-(\omega-\omega_0)^2/4\sigma^2)$, and duration $\sim 1/\sigma$. The total number of photons is $N = \int |\alpha(\omega)|^2 d\omega = |\alpha_0|^2$. And we find $K_{coh}(x) = |\alpha_0|^2 \exp(-x^2/8\sigma^2)$. Then, for two-photon resonance, $2\omega_0 = \omega_{fg}$, the excitation probability is:

$$P_f^{coh} = |\alpha_0|^4 \Sigma^{(2)} L_0^4 \xi(\gamma_{fg}/2\sigma) \tag{25}$$

where $\xi(z) = \exp(+z^2) erfc(z)$, (sometimes denoted as $erfcx(z)$) which has maximum value 1 at $z = 0$ (an ultrashort pulse). For a long, quasi-monochromatic pulse (large z), $\xi(z)$ decays to zero as $\sim 1/(\pi^{1/2} z)$. We summarize these two limits:

$$P_f^{coh} = \begin{cases} |\alpha_0|^4 \Sigma^{(2)} L_0^4 \dfrac{1}{\sqrt{\pi}} \dfrac{2\sigma}{\gamma_{fg}} & (\gamma_{fg} \gg \sigma) \\ |\alpha_0|^4 \Sigma^{(2)} L_0^4 & (\gamma_{fg} \ll \sigma) \end{cases} \quad (26)$$

We see, again, the expected intensity-squared dependence. Importantly, in the impulsive limit $\sigma \gg \gamma_{fg}$ the probability does not depend on $\gamma_{fg}$ or $\sigma$ because for the nonlinear TPA process the effect of spreading the spectrum over a broader range is compensated by the increasing peak intensity in the time domain, as verified in **Appendix C** also for an arbitrary pulse shape.

### 4.2. Coherent state TPA, arbitrary long pulse

In the case of a coherent state in a long, quasi-monochromatic pulse with arbitrary shape, Eq.(23) leads to the population $P_f = 2\operatorname{Re}[p_f]$ (**Appendix D**):

$$P_f^{coh} = 2\Sigma^{(2)} L_0^4 \frac{\gamma_{fg}}{\gamma_{fg}^2 + (\omega_{fg} - 2\omega_0)^2} \int |A(t)|^4 dt \quad (27)$$

where $A(t)$ is the time-domain envelope defined in Eq., where $|A(t)|^2$ is in units of photons per second. For exact resonance, $\omega_P = 2\omega_0 = \omega_{fg}$, the relation between Eq.(27) and the upper line of (26) is seen by noting that for a constant (rectangular) pulse of duration $T$, $\int |A(t)|^4 dt = N^2/T = |\alpha_0|^4/T$, and for the Gaussian pulse, $\int |A(t)|^4 dt = |\alpha_0|^4 \sigma/\sqrt{\pi}$ thus showing a correspondence $T \sim \sqrt{\pi}/\sigma$, the prefactor being a consequence of the nonlinear response in TPA.

### 5. Two-photon absorption (TPA) with isolated EPP

We now come to the heart of the matter—TPA with entangled photon pairs. For the state $|\Psi\rangle$, Eq.(19), created by Type-I (or Type-0) SPDC into a single spatial-and-polarization mode, the correlation function in Eq.(12) is evaluated, denoting $\tilde{\omega}' \equiv \omega + \tilde{\omega} - \omega'$, as:

$$C_\omega^{(4)} = \langle \Psi | \hat{a}^\dagger(\omega') \hat{a}^\dagger(\tilde{\omega}') \hat{a}(\omega) \hat{a}(\tilde{\omega}) | \Psi \rangle \qquad (28)$$
$$= \varphi^*(\omega',\tilde{\omega}') \varphi(\omega,\tilde{\omega})$$

where:

$$\varphi(\omega,\tilde{\omega}) = \langle vac | \hat{a}(\omega) \hat{a}(\tilde{\omega}) | \Psi \rangle \qquad (29)$$
$$= \varepsilon \langle vac | \hat{a}(\omega) \hat{a}(\tilde{\omega}) \int d\omega_1 \int d\omega_2 \, \psi(\omega_1,\omega_2) \hat{a}^\dagger(\omega_1) \hat{a}^\dagger(\omega_2) | vac \rangle$$

Using the symmetry $\psi(\omega,\tilde{\omega}) = \psi(\tilde{\omega},\omega)$ and the relation:

$$\langle vac | \hat{a}(\omega) \hat{a}(\tilde{\omega}) \hat{a}^\dagger(\omega_1) \hat{a}^\dagger(\omega_2) | vac \rangle \qquad (30)$$
$$= 2\pi\delta(\omega-\omega_1) 2\pi\delta(\tilde{\omega}-\omega_2) + 2\pi\delta(\omega-\omega_2) 2\pi\delta(\tilde{\omega}-\omega_1)$$

we find:

$$\varphi(\omega,\tilde{\omega}) = 2\varepsilon\psi(\omega,\tilde{\omega}) \qquad (31)$$

which is the quantum amplitude for two photons of frequencies $(\omega,\tilde{\omega})$ to participate in the interaction. Then:

$$p_f = \varepsilon^2 4 \Sigma^{(2)} L_0^4 \int d\omega \int d\tilde{\omega} \int d\omega' \, \frac{\psi^*(\omega',\omega+\tilde{\omega}-\omega')\psi(\omega,\tilde{\omega})}{\gamma_{fg} - i\omega_{fg} + i(\omega+\tilde{\omega})} \qquad (32)$$

For Type-II SPDC, where the signal and idler modes are distinct, Eq.(32) still holds, with the replacement: $\psi(\omega,\tilde{\omega}) \to \Psi(\omega,\tilde{\omega}) = \{\psi(\omega,\tilde{\omega}) + \psi(\tilde{\omega},\omega)\}/2$. Thus, even though $\psi(\omega,\tilde{\omega})$ need not be symmetric, $\Psi(\omega,\tilde{\omega})$ is symmetric. See **Appendix E** for proof.

The arguments of the quantum amplitudes appearing in the numerator are consistent with the fact that when two photons are absorbed over a lengthy time interval, their energies must sum to a particular value, as in the double-sided Feynman diagram in **Fig. 1c**. In particular, the two arguments of the function $\psi$ sum to $\omega+\tilde{\omega}$ in both instances.

Eq.(32) is of the same form as the coherent-state result Eq.(22) with the replacement $\alpha(\omega)\alpha(\tilde{\omega}) \to 2\varepsilon\psi(\omega,\tilde{\omega})$, that is, the EPP result resembles the coherent-state result but with a generalized spectral dependence: the coherent state takes the form of a separable two-photon state without frequency correlations. And the two have different photon-flux dependences: linear ($\varepsilon^2$) for the EPP state and quadratic ($|\alpha|^4$) for the coherent state, as expected. (Frequency correlations could be built into the 'classical' model using a statistical mixture of coherent states, as in [31], but Schlawin and Buchleitner argue that such states do not enhance the absorption probability above the pure-state case, [18] and Lerch and Stefanov show that a statistical mixture of correlated monochromatic states can mimic the frequency correlations but not the time correlations. [32])

Changing variables to $x = (\omega + \tilde{\omega} - 2\omega_0); z = \omega - \omega_0; z' = \omega' - \omega_0$, as for the coherent-state case, we obtain its generalization:

$$p_f = \varepsilon^2 4 \Sigma^{(2)} L_0^4 \int dx \int dz \int dz' \frac{\psi(\omega_0 + z, \omega_0 + x - z)\psi^*(\omega_0 + z', \omega_0 + x - z')}{\gamma_{fg} - i\omega_{fg} + i(2\omega_0 + x)} \quad (33)$$

Here the sum of the arguments of the function $\psi$ sum to $2\omega_0 + x$, which is the same as appears in the denominator. Thus, for a given detuning $x$ from two-photon resonance, we interpret $\psi(\omega_0 + z, \omega_0 + x - z)$ and $\psi^*(\omega_0 + z', \omega_0 + x - z')$ as the two-photon amplitudes that act on the two sides of the double-sided Feynman diagram in **Fig. 1**. The integration over $z$ reflects that different combinations of frequencies can sum to $2\omega_0 + x$.

We can write the result more compactly as:

$$p_f = \varepsilon^2 4 \Sigma^{(2)} L_0^4 \int dx \frac{|K_\psi(x)|^2}{\gamma_{fg} - i\omega_{fg} + i(2\omega_0 + x)} \quad (34)$$

with

$$K_\psi(x) = \int dz\, \psi(\omega_0 + z, \omega_0 + x - z) \quad (35)$$

$K_\psi(x)$ represents the integrated amplitude for excitation at a particular two-photon detuning, as explained in detail in **Section 6**.

The general case as given here can be analyzed quantitatively by direct numerical integration. Here, instead, we consider two important special cases that can be treated analytically. In **Appendix F** the case of an ultrashort EPP pulse in the impulsive limit is treated.

### 5.1 EPP with long pump pulse

An important example of time-frequency entanglement is that of EPP generated by SPDC using a narrow-band pump pulse with long duration $T$. Energy conservation localizes the JSA $\psi(\omega, \tilde{\omega})$ along the antidiagonal, $\tilde{\omega} + \omega = \omega_P$. For degenerate SPDC the JSA has a single peak at the frequency $\omega_0 = \omega_P / 2$. In the absence of dispersion, which creates phase correlations, we can model the JSA as the product of (square-normalized) narrow and broad functions, $\psi_N(\omega)$ and $\psi_B(\omega)$, centered at $\omega = \tilde{\omega} = \omega_P / 2$ and oriented along diagonal and antidiagonal axes in the $(\omega, \tilde{\omega})$ plane, respectively. The width of $\psi_N(\omega)$ is the linewidth (inverse duration) of the pump pulse, and the linewidth of $\psi_B(\omega)$ is $\sqrt{2}$ times the spectral width of the EPP, set by the phase-matching conditions. Then:

$$\psi(\omega,\tilde{\omega}) = \psi_B(\frac{\omega-\tilde{\omega}}{2})\psi_N(\omega+\tilde{\omega}-\omega_P) \qquad (36)$$

where, by state symmetry, $\psi_B(-x) = \psi_B(x)$, and both functions are square-normalized in $dx/2\pi$. Then we find easily:

$$K_\psi(x) = \psi_N(x)\int dz\, \psi_B(z) \qquad (37)$$

This gives the complex probability:

$$p_f = \varepsilon^2 4\Sigma^{(2)} L_0^{\,4} \left|\int dz\, \psi_B(z)\right|^2 \int dx \frac{|\psi_N(x)|^2}{\gamma_{fg} - i\omega_{fg} + i(2\omega_0 + x)} \qquad (38)$$

Equation (38) shows clearly that it is the narrow function $\psi_N(x)$ that effectively drives the molecular transition, as a result of the anticorrelation of EPP frequencies.

The integral can be evaluated similarly to the coherent-state case Eq.(25) by assuming the Gaussian forms (valid for Type-I SPDC only in the case of a long narrow-band pump pulse):

$$\begin{aligned}\psi_N(x) &= (\sigma_N^{\,2}/2\pi)^{-1/4} \exp(-x^2/4\sigma_N^{\,2}) \\ \psi_B(x) &= (\sigma_B^{\,2}/2\pi)^{-1/4} \exp(-x^2/4\sigma_B^{\,2})\end{aligned} \qquad (39)$$

where we must impose the condition $\sigma_N < \sigma_B$. The 'narrow' width $\sigma_N$ equals the spectral width of the laser pulse driving the SPDC, while the 'broad' width $\sigma_B$ is determined by phase matching. [33]  In this case,

$$K_\psi(x) = \frac{2\sigma_B}{\sigma_N}\exp[-x^2/(2\sigma_N^{\,2})] \qquad (40)$$

This leads to the probability:

$$P_f^{EPP} = \varepsilon^2 4\Sigma^{(2)} L_0^{\,4} \frac{2\sigma_B}{\sigma_N}\xi\left(\gamma_{fg}/\sqrt{2}\sigma_N\right) \qquad (41)$$

Where again $\xi(z) = \exp(+z^2)\mathit{erfc}(z)$. A related expression was given in [34]. In two limits this becomes:

$$P_f^{EPP} = \begin{cases} \varepsilon^2 4\Sigma^{(2)} L_0^4 \dfrac{2}{\sqrt{\pi}} \dfrac{\sqrt{2}\sigma_B}{\gamma_{fg}} & (\gamma_{fg} \gg \sigma_N) \\ \varepsilon^2 4\Sigma^{(2)} L_0^4 \dfrac{2\sigma_B}{\sigma_N} & (\gamma_{fg} \ll \sigma_N) \end{cases} \qquad (42)$$

The second of these two expressions, for $\gamma_{fg} \ll \sigma_N$, is in the impulsive limit with respect to the correlation duration ($\sim 1/\sigma_B$) of the EPP wave packet. (Compare with the second line of Eq.(26).) In this Gaussian model of the JSA, the ratio $\sigma_B/\sigma_N$ is a measure of the number of temporal (time-frequency) modes in the EPP state that impinge on the molecule in a single pulse with duration $\sim 1/\sigma_N$. [35] Given that the EPP always arrive together within a time $1/\sigma_B$, regardless of the duration ($1/\sigma_N$) of the pulse, the probability for TPA is enhanced by this factor relative to a narrow-band coherent pulse of the same duration, wherein the photons arrive independently.

The first of the two expressions in Eq.(42) can be interpreted as the number of temporal modes in the EPP state that impinge on the molecule in the time $1/\gamma_{fg}$, the molecular coherence time, which in this case is much smaller than the EPP pulse duration $(1/\gamma_{fg} \ll 1/\sigma_N)$.

We note that the Gaussian approximation can also be applied to Type-II SPDC, with similar conclusions for the long-pulse case.

### 5.2 Effect of dispersion on TPA by EPP

The frequency-dependent refractive index of optical elements that light passes through before reaching the sample is known to temporally broaden ultrashort pulses and reduce their effectiveness in nonlinear optical processes. In EPP-driven TPA, the effects of dispersion are expected to be similar. To account for such effects, we incorporate dispersive propagation into the two-photon JSA by replacing:

$$\psi(\omega,\tilde{\omega}) \to \psi(\omega,\tilde{\omega}) \exp[i(D/2)(\omega^2 + \tilde{\omega}^2)] \qquad (43)$$

Where $D$ is the second-order (group delay) dispersion of the transmitting optical system. Inserting this expression into Eq.(35) and using the Gaussian forms in Eq.(39) (valid for long SPDC pump) leads to:

$$K_\psi(x) = \dfrac{2\sigma_B \exp[-x^2/(2\sigma_N^2)]}{\sigma_N\sqrt{1 + 16 D^2 \sigma_B^4}} \qquad (44)$$

Comparing to Eq.(40), we see the sole effect of second-order dispersion is to replace $\sigma_B$ by

$$\tilde{\sigma}_B = \frac{\sigma_B}{\sqrt{1+16 D^2 \sigma_B^4}} \tag{45}$$

which means the EPP correlation (entanglement) time $\tau_{corr} \sim \sigma_B^{-1}$ is increased according to:

$$\tau_{corr} \to \tau_{corr}\sqrt{1+16 D^2 \sigma_B^4} \tag{46}$$

The $x$ dependence of $K_\psi(x)$ is not altered; there is only an overall decrease of magnitude of $K_\psi(x)$ and thus a decrease of the TPA probability.

## 5.3 EPP with separable (factorable) JSA

As a special case, if an ultrashort pump pulse together with a particular phase-matching condition of the SPDC create a separable (factorable) JSA, as in [33], then because of symmetry, $\psi(\omega,\tilde{\omega}) = \phi_0(\omega)\phi_0(\tilde{\omega})$. Then:

$$p_f = \varepsilon^2 4 \Sigma^{(2)} L_0^4 \int dx \frac{|K_0(x)|^2}{\gamma_{fg} - i\omega_{fg} + i(2\omega_0 + x)} \tag{47}$$

where $K_0(x)$ is given by

$$K_0(x) = \int dz\, \phi_0(\omega_0 + z)\phi_0(\omega_0 + x - z) \tag{48}$$

This expression has precisely the same spectral dependence as for the coherent-state pulse in Eq.**(23)**. In this case there is no spectral entanglement, and thus no enhancement of TPA through spectral correlation.

## 6. TPA enhancement by number correlation and spectral correlation

In all of the EPP examples discussed the excitation probability has the form:

$$P_f^{EPP} = \varepsilon^2 \times \eta_{EPP} \tag{49}$$

$\varepsilon^2$ is the probability that in the time window $T$ there exists one photon pair, and $\eta_{EPP}$ is the conditional probability that, given a pair is present, the molecule will be excited. So, we can express this as:

$$P_f^{EPP} = Prob(pair) \times Prob(excitation\,|\,pair) \tag{50}$$

Recall it is assumed that *Prob*(*pair*) is significantly less than 1, in order to avoid multiple pairs from the SPDC.

A fruitful way to compare the EPP results with the coherent-state results, is to consider cases where the mean number of coherent-state photons, $N = |\alpha_0|^2$, equals the mean number of EPP photons, $2\varepsilon^2$, or equivalently $|\alpha_0|^2/2 = \varepsilon^2$.

In comparison, for a coherent state, the photon statistics are Poisson, and for mean number much less than one, the probability for one photon equals the mean number $|\alpha_0|^2$, and the probability for two photons equals $|\alpha_0|^4/2$. Therefore, in this equal-flux regime, we can write all the coherent-state results in a form analogous to the EPP results in Eq.(49), as:

$$P_f^{coh} = (|\alpha_0|^4/2) \times \eta_{coh} \tag{51}$$

and interpret it in the same manner as for Eq.(50).

We refer to $\eta_{EPP}$ and $\eta_{coh}$ as the 'spectral efficiencies' of the various excitation processes. The separation into photon-pair probabilities and spectral efficiencies allows us to examine the two contributors to TPA enhancement by EPP—the photon number correlations and the spectral correlations. Both are aspects of quantum entanglement in the EPP state.

A useful quantity for comparing excitation scenarios is the 'quantum enhancement factor' $QEF$, defined as the ratio of EPP excitation probability to coherent-state excitation probability for a common interaction time $T$:

$$QEF = \frac{P_f^{EPP}}{P_f^{coh}} = QEF_{number} \times QEF_{spectral} \tag{52}$$

where

$$QEF_{number} = \frac{\varepsilon^2}{|\alpha_0|^4/2} \tag{53}$$

is the enhancement due to photon-number correlation and

$$QEF_{spectral} = \frac{\eta_{EPP}}{\eta_{coh}} \tag{54}$$

is the enhancement due to spectral correlation.

The spectral part of the QEF for the case of TPA with *resonant* intermediate states was considered in detail in [18]. We treat only the case with non-resonant intermediate states, and we discuss both aspects of enhancement—number and spectral correlations.

First, note that, as defined, the enhancement due to number correlation is the same regardless of pulse durations. It depends only on the fact that for EPP, a photon is always accompanied by one other, whereas for a coherent state one relies on an 'accidental' pairing of photons within the interaction time $T$, which is assumed to be long enough that the interaction proceeds to completion—creating an excited state or not. Therefore, Eq.(53) gives $QEF_{number}$ in all cases. The key point is that $QEF_{number}$ becomes very large only in the regime that an extremely small mean number of photons are present in the interaction time $T$. The early experiments by Georgiades et al on TPA of atomic cesium observed such enhancement in this regime approaching an order of magnitude. [11]

Second, consider enhancement due to spectral correlation of EPP. The general form of the spectral enhancement factor is, by taking the real parts of Eqs. (23) and (34):

$$QEF_{spectral} = \frac{\int dx \frac{|K_\psi(x)|^2}{\gamma_{fg}^2 + (\omega_{fg} - 2\omega_0 - x)^2}}{\int dx \frac{|K_{coh}(x)|^2}{\gamma_{fg}^2 + (\omega_{fg} - 2\omega_0 - x)^2}} \tag{55}$$

where $K_\psi(x)$ and $K_{coh}(x)$ are given for EPP and a coherent state by Eqs.(35) and (24). We see that if $|K_\psi(x)| = |K_{coh}(x)|$, then $QEF_{spectral} = 1$ and there is no spectral enhancement by EPP. (There is still a relative enhancement by number correlation if the mean photon number per pulse is much less than one.)

To understand the meaning of this formula generally, recall that $K_\psi(x) = \int dz\, \psi(\omega_0 + z, \omega_0 + x - z)$ is the integrated amplitude for absorption at a particular two-photon detuning $x$. This statement is illustrated geometrically in **Fig. 3**, which shows that $K_\psi(x)$ is the *diagonal* projection of $\psi(\omega, \tilde{\omega})$ onto the $\omega$ axis, creating a distribution centered around $2\omega_0$. The same procedure holds for the diagonal projection of the coherent-state two-photon amplitude, which is the product $\phi(\omega)\phi(\tilde{\omega})$. We then define the 'diagonally-projected spectra' as the mod-squares, $|K_\psi(x)|^2$ and $|K_{coh}(x)|^2$.

To aid in fair comparison of EPP and coherent states, we also define the 'marginal spectrum' of the EPP state as the mod-squared projected *vertically* onto the $\omega$ axis:

$$M_\psi(\omega) = \int d\tilde{\omega}\, |\psi(\omega, \tilde{\omega})|^2 \tag{56}$$

which for the Gaussian model in Eq.(39) is proportional to $\exp(-(\omega-\omega_0)^2/2\sigma_B^2)$ for $\sigma_N \ll \sigma_B$. Likewise define a marginal spectrum for the coherent state:

$$M_{coh}(\omega) = \int d\tilde{\omega} |\phi(\omega)\phi(\tilde{\omega})|^2 \qquad (57)$$
$$= |\phi(\omega)|^2$$

which is identical to the standard spectrum for the coherent state. The marginal spectrum corresponds to the spectrum as measured by a conventional grating spectrometer with linear-response detector, whereas the diagonally projected spectrum is that 'felt' by the molecule, which acts as a spectrally selective two-photon detector. Specific examples are treated below.

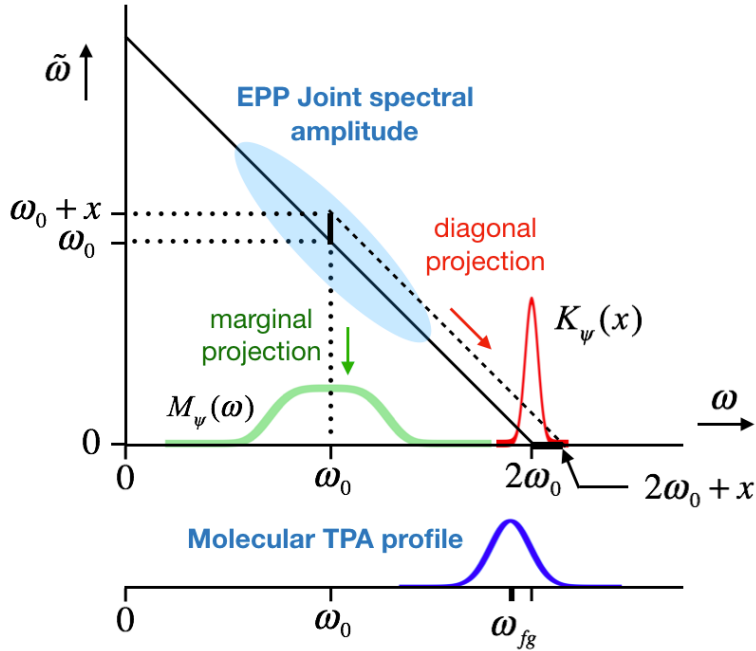

Fig. 3 The marginal (vertical) and diagonal projections of the JSA, $|\psi(\omega,\tilde{\omega})|$, of a two-photon state of the field may be broader or narrower than two-photon absorption profile.

Our goal is to quantify the enhancements to TPA by EPP relative to coherent states. First consider cases where EPP are generated by SPDC using a long narrow-band Gaussian pump pulse of duration $T \sim 1/\sigma$, as illustrated in **Fig.2** and treated in **Sec. 5.1**. Energy conservation localizes the JSA $\psi(\omega,\tilde{\omega})$ along the antidiagonal, $\tilde{\omega}+\omega=\omega_P$, as shown in **Fig. 3**. The diagonally projected amplitude, which is responsible for TPA, can be far narrower than the marginal spectrum. Then, EPP can be far more efficient spectrally than a coherent state with the same marginal spectral width, as discussed following Eq.(42). To elaborate, the spectral enhancement factor in the absence of dispersion is, from Eqs.(25) and (41):

$$QEF_{spectral} = \frac{\eta_{EPP}}{\eta_{coh}} = \frac{2\sigma_B}{\sigma_N} \frac{\xi(\gamma_{fg}/\sqrt{2}\sigma_N)}{\xi(\gamma_{fg}/2\sigma)} \tag{58}$$

where it is understood that $\sigma_N < \sigma_B$. This expression has the four limiting values:

$$QEF_{spectral} = \begin{cases} \dfrac{\sqrt{2}\sigma_B}{\sigma} & (\gamma_{fg} \gg \sigma, \sigma_N) \\[6pt] \dfrac{2\sigma_B}{\sigma_N} & (\gamma_{fg} \ll \sigma, \sigma_N) \\[6pt] \dfrac{2\sqrt{2}}{\sqrt{\pi}} \dfrac{\sigma_B}{\gamma_{fg}} & (\gamma_{fg} \gg \sigma_N)(\gamma_{fg} \ll \sigma) \\[6pt] \dfrac{\sqrt{\pi}\sigma_B}{\sigma_N} \dfrac{\gamma_{fg}}{\sigma} & (\gamma_{fg} \ll \sigma_N)(\gamma_{fg} \gg \sigma) \end{cases} \tag{59}$$

Recall that the spectral full width in rad/s of the coherent-state pulse is $2\sigma$ while the full width of the EPP marginal spectrum is $\sqrt{2}\sigma_B$. And $2\sigma_N$ is the bandwidth of the pump laser driving the EPP source.

The first of these limits corresponds to a very broad two-photon absorption line and compares excitation by a narrow-band coherent state to excitation by EPP pumped by a narrow-band laser. The EPP spectral width $\sqrt{2}\sigma_B$ may be small or large. If the EPP and coherent state have roughly the same (marginal) spectral widths, this QEF has value near 1 and there is no spectral enhancement. But if $\sqrt{2}\sigma_B \gg \sigma$, then spectral enhancement can be quite large in this comparison, as discussed in the Introduction.

The second of these limits corresponds to a relatively narrow two-photon absorption line and compares excitation by a short broad-band coherent pulse to excitation by EPP pumped by a short broad-band laser pulse. Again, spectral enhancement can be large if $\sqrt{2}\sigma_B \gg \sigma$ and dispersion is well corrected. This result is analogous to that in Eq.3 of [12].

The third and fourth limits correspond to mixed scenarios, where only one of the coherent-state pulse or the pulse driving the EPP source is narrower than the absorption line.

The effect of dispersion on these conclusions is easy to state. We showed in **Section 5.2** that the sole effect of second-order dispersion is to replace $\sigma_B$ by $\tilde{\sigma}_B = \sigma_B / \sqrt{1 + 16 D^2 \sigma_B^4}$. The shapes of $|\psi(\omega,\tilde{\omega})|$ and $K_\psi(x)$ are not altered by dispersion, so the projection pictures in **Fig. 3** still apply. The TPA probability is merely decreased by the factor $1/\sqrt{1+16 D^2 \sigma_B^4}$.

Now, as in **Fig. 4**, consider cases where the JSA of the EPP takes on one of the possible more general forms.

**Figure 4a** illustrates the case of separable JSAs. For coherent states, the JSA is necessarily separable and circular-shaped, as shown. For EPP, the JSA can be made separable by using special phase-matching design. [30] As stated earlier, if both are separable, and have the same marginal-spectrum width, then they also have similar diagonally projected widths, and thus no significant spectral enhancement is present.

**Figure 4b** illustrates the case of non-separable JSA for EPP, with near-circular support. This form of JSA is typical when pumping Type-I SPDC with sub-ps pulses; the JSA is nonzero in a roughly circular region, but its shape may be, for example, crescent-like, as shown. If that region has roughly the same width as a comparator coherent state, then, again, no significant spectral enhancement is possible.

**Figure 4c** illustrates the case of a non-separable JSA for EPP, elongated on the diagonal. Although not easily generated using SPDC in a second-order nonlinear-optical crystal, such a JSA is possible using spontaneous four-wave mixing in a third-order nonlinear-optical medium, such as optical fiber. In this case, the EPP has a broader spectrum when projected diagonally, and would be *less* effective in TPA than a coherent state having the same width of its marginal spectrum. (Such a *positive* frequency correlation can lead to enhancement of stimulated Raman transitions, as demonstrated with classically correlated fields in [36].

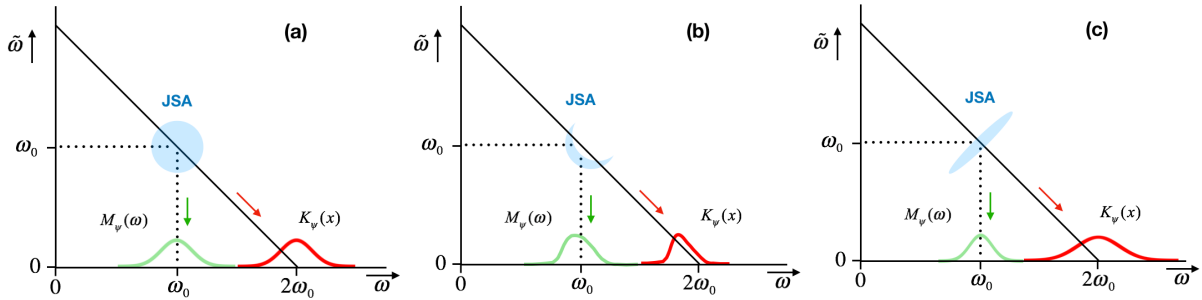

Fig. 4 Example JSAs showing the vertical and diagonal projections of a two-photon state.

## 7. TPA cross section, optical fluxes and event rates

In experiments, a commonly measured quantity is the mean flux of the light used for exciting the molecular sample. Here we analyze our results in terms of optical fluxes for coherent states and for EPP. Because the probability for detecting TPA-fluorescence is small, photon counting is used in experiments. In such experiments, the observed signal is the count rate $R_{count}$, the number of TPA-induced fluorescent photons detected per second. Here we assume for simplicity that every TPA event leads to a fluorescent photon (quantum efficiency equals one), and that a combined collection and detection efficiency $\eta_{det}$ is common to all considered experiments.

We first give the expression for the 'conventional' two-photon absorption (excitation) cross section $\sigma^{(2)}$, driven at two-photon resonance by a monochromatic light source, in terms of our formalism (see **Appendix G**):

$$\sigma^{(2)} = 2\Sigma^{(2)} L_0^4 \frac{A_0^2}{\gamma_{fg}} \tag{60}$$

An approximate evaluation of $\sigma^{(2)}$ yields on the order of $10^{-49} cm^4 s / photon^2$, in agreement with known values for typical dye molecules. [37, 38] Dendrimers have shown values as high as $10^{-46} cm^4 s / photon^2$, [39, 38] so we take this to be a reasonable upper bound value for molecules.

All of the previously given results can be rewritten in terms of the cross section. For example, denote the coherent-state mean flux (photons s$^{-1}$) as $F_{coh} = |A(t)|^2$, so the mean number of photons delivered in the entire time window of duration $T$ is $N = \int F_{coh}(t) dt = |\alpha_0|^2$. Then the $f$-state probability for a single molecule driven by a constant, resonant quasi-monochromatic coherent state for an interaction time $T$, from Eq.(27), is the expected form:

$$\begin{aligned} P_f^{coh} &= \frac{\sigma^{(2)}}{A_0^2} \int |A(t)|^4 dt \\ &= \sigma^{(2)} \left( \frac{F_{coh}}{A_0} \right)^2 T \end{aligned} \tag{61}$$

In defining mean flux for EPP a subtlety arises, in that the perturbation theory used above for the EPP case assumes that either zero or two photons are present in the exciting field during the time window $T$ defining the interaction between light and molecule. This condition defines the 'isolated EPP' regime. Therefore, we require that the interaction time window be longer than the EPP correlation time (inverse EPP bandwidth), and that the pulse must be longer than the molecular dephasing time $1/\gamma_{fg}$ in order that a photon pair interacts completely with the molecule before a second photon pair arrives. In the case of CW excitation, we imagine the CW laser pumping the SPDC to be made of a series of rectangular pulses with constant power and duration $T$, as shown earlier in **Fig.2(a)**. When considering TPA with short pulses of SPDC or coherent-state light, it suffices to consider a single pulse occupying the same interaction time window $T$, as in **Figs.2(b) and (c)**. These considerations reveal the mean photon flux (twice the pairs flux), time-averaged over the whole interval $T$, to be $F_{EPP} = 2\varepsilon^2 / T$.

The $f$-state probability for a molecule driven by a constant EPP flux for a long interaction time $T$ is, then, from Eq.(42):

$$P_f^{EPP} = \begin{cases} 2F_{EPP}T\dfrac{\sigma^{(2)}}{A_0^2}\dfrac{\sqrt{2}}{\sqrt{\pi}}\sigma_B & (\gamma_{fg} \gg \sigma_N) \\ 2F_{EPP}T\dfrac{\sigma^{(2)}}{A_0^2}\dfrac{\gamma_{fg}\sigma_B}{\sigma_N} & (\gamma_{fg} \ll \sigma_N) \end{cases} \qquad (62)$$

To compare the long-pulse EPP excitation probability, Eq.(62), to narrow-band coherent-state excitation, Eq.(61), we rewrite the quantum enhancement factor, Eq.(52), using Eq.(59) and $|\alpha_0|^2 = F_{coh}T$, considering cases of large or small $\gamma_{fg}$:

$$QEF = \dfrac{F_{EPP}}{F_{coh}^2} \times \begin{cases} \dfrac{\sqrt{2}\sigma_B}{T\sigma} & (\gamma_{fg} \gg \sigma, \sigma_N) \\ \dfrac{\sqrt{2}\sigma_B}{T\sigma_N/\sqrt{2}} & (\gamma_{fg} \ll \sigma, \sigma_N) \end{cases} \qquad (63)$$

where it is understood that $\sigma_N < \sigma_B$. The top line of Eq.(63) can be summarized as:

$$QEF \simeq \dfrac{F_{EPP}B\chi}{F_{coh}^2} \qquad (64)$$

where we denoted the (marginal) bandwidth of EPP as $B = 2\sigma_B/\sqrt{2} = \sqrt{2}\sigma_B$, and we define $\chi = \sqrt{2}/T\sigma_N$ as a factor that has value of order unity. For equal mean fluxes, denoted by $F_{EQ}$, where $F_{EQ} = F_{coh} = F_{EPP} = 2\varepsilon^2/T$, this gives:

$$QEF = \dfrac{B\chi}{F_{EQ}} = \dfrac{BT\chi}{2\varepsilon^2} \qquad (65)$$

which is a major result and is stated in the Introduction. Therefore, for low enough flux that there are no overlapping EPPs, the overall TPA enhancement can, in principle, be large. The time-bandwidth product $BT$ is a measure of the number of temporal (time-frequency) modes that impinge on the molecule in time $T$, and can be large. And the denominator $\varepsilon^2$ is the probability ($< 1$) that an EPP is present in the long time interval $T$. Such a pair of entangled photons is spread over this interval (consistent with the assumption we made in the perturbation theory), and these photons are tightly correlated in time (to within roughly $1/B$).

A typical example is a 1-ns pulse—long on relevant times scales for molecular absorption in solution—where the absorption full linewidth in Hz, $2\gamma_{fg}/2\pi$, is of the order of 10 THz. If the EPP has Gaussian marginal spectrum with bandwidth $B = 10\ THz$, and the probability of a pair is $\varepsilon^2 = 0.1$, then $QEF$, is of the order of $10^5$, rather large. On the other hand, in this regime of large enhancement the absolute rate of TPA is extremely small, making it difficult, if not impossible, to observe experimentally, as now discussed.

The rate of TPA for a single molecule illuminated by narrow-band, two-photon-resonant light coherent-state light is:

$$Rate_{coh} = \sigma^{(2)} \left(\frac{F_{coh}}{A_0}\right)^2 \qquad (66)$$

As an example, consider a 1-cm thick dye sample (R6G) with concentration 2 millimolar (i.e., $1.2 \times 10^{18}\ cm^{-3}$). The TPA cross section at wavelength 1064 nm is about 9 GM (where 1 GM = $10^{-50}\ cm^4 s/photon^2$).[40] Illuminating with a 1064-nm CW laser with beam area $10^{-6}$ cm², and power 100 mW, yields a TPA absorption rate $3 \times 10^{10}/s$, and assuming a collection-plus-detection efficiency of 0.01, predicts a total TPA fluorescence count rate of about $3 \times 10^8/s$, in reasonable agreement with typical experiments

Now consider the corresponding rate for EPP illumination. We can express the optimally enhanced $f$-state probability after one constant-flux EPP pulse of duration $T$ as:

$$P_f^{EPP} = Rate_{EPP} T \qquad (67)$$

Where the enhanced rate is:

$$Rate_{EPP} = Rate_{coh} \cdot QEF$$
$$= \left(\sigma^{(2)} \frac{B\chi}{A_0}\right)\frac{F_{EPP}}{A_0} \qquad (68)$$

We have written Eq.(68) in a form allowing us to identify the term in brackets as the so-called entangled TPA cross section $\sigma_e^{(2)}$, with the same units, $cm^2$, as a one-photon absorption cross section. This form is analogous to that introduced phenomenologically by Fei et. al. in [19] as $\sigma_e^{(2)} = \sigma^{(2)}/2A_e T_e$, where $A_e$ is called the entanglement area (area within which EPP are spatially correlated) and $T_e$ is the entanglement time (interval within which EPP are temporally correlated). In our case (and in most experiments) the photons within an EPP are independently spread within a single-mode transverse beam profile, so $A_e = A_0$, and $T_e = 1/B$. Recall the factor $\chi$ is of order unity, so that our rigorous derivation reproduces the phenomenological one in [19].

But, crucially, our result contradicts that given by Saleh et. al. in their Eq.1. [5] To explain this claim, we generalize Eq. (68) to the previously discussed case of a Gaussian JSA, using Eqs.(66), (58) and (53), and the correspondence $T \sim \sqrt{\pi}/\sigma$ mentioned after Eq.(27), to give:

$$Rate_{EPP} = \left( \frac{\sigma^{(2)}}{A_0} \frac{2\sigma_B \sigma}{\sqrt{\pi}\sigma_N} \frac{\xi(\gamma_{fg}/\sqrt{2}\sigma_N)}{\xi(\gamma_{fg}/2\sigma)} \right) \frac{F_{EPP}}{A_0} \qquad (69)$$

Again, the term in brackets is the entangled TPA cross section $\sigma_e^{(2)}$. It replaces that given by Saleh et. al. in their Eq.1 [5], which uses the model of a sinc-function EPP spectrum, which has (physically unrealistic) long tails and therefore overestimates the effects of intermediate-state resonances, according to de Leon-Montiel et. al. [34]

For the R6G dye experiment just mentioned, illuminating the same sample with a CW 1064-nm EPP beam in the isolated-pair regime with power 20 nW ($10^{11}$ $photons/s$) and bandwidth $1 \times 10^{13}$ $Hz$, yields a quantum-enhanced TPA rate of $0.1$ $events/s$, and assuming a collection-plus-detection efficiency of 0.01, predicts a TPA fluorescence count rate of about $1 \times 10^{-3}/s$. Without much greater enhancement than that predicted by the present theory, such a fluorescence signal is unobservable, consistent with our recent experiments. [15]

The use of a solution of a dendrimer with TPA cross section 10,000 GM, in a similar experiment would not lead to a different qualitative conclusion.

In principle (although unknown in practice), the rate of EPP generation from the SPDC source could be increased several orders of magnitude to about $10^{12}$ $pairs/s$, while remaining in the isolated-EPP regime, but even then, the detected TPA fluorescence count rate would not be easily detectable.

As an alternative way to understand the prediction, we define the 'cross-over point' as the value of photon flux at which the coherent-state TPA rate equals the EPP-induced TPA rate. Equation (64), with $QEF = 1$ and $F_{cross} = F_{EPP}$, yields the cross-over value $F_{cross} = B\chi$. With $B = 10^{13} s^{-1}$ and $\chi = 1$, as before, the cross over occurs at $F_{cross} = 2 \times 10^{13}$ $photons/s$.

The observation of beam attenuation (observable in transmission) by isolated EPP two-photon absorption is similarly challenging. With an incident EPP flux of $10^{11}$ $pairs/s$ and a TPA event rate in the full volume of around $10^{-5}/s$ to $10^{-2}/s$, the fractional absorption is minuscule.

The effect of dispersion (**Section 5.2**) is to decrease the TPA probability by the factor $1/(1 + 16 D^2 \sigma_B^4)^{1/2}$. A typical value of D for a 1-m path in fused silica is 50,000 fs², leading to a predicted decrease in TPA by around 1/120 for a marginal EPP bandwidth of 30 nm ($\sigma_B = 2.5 \times 10^{13}$ $rad/s$) at 1064 nm wavelength.

## 5. Discussion and Conclusions

The bottom-line conclusion of this study is the prediction that, for realistic experimental scenarios, the regime in which quantum enhancement of TPA in typical molecules by photon-number correlations and spectral correlation in the isolated EPP regime is significant corresponds to such low TPA event rates as to be unobservable in practice. On a more positive note, this study clarifies qualitatively and quantitatively the origins of such enhancements, which are predicted to be achievable in simpler systems such as gas-phase atoms and possibly color centers (defects) in semiconductors, which have narrow transitions and large dipole matrix elements. [12]

In particular, we showed that the enhancement due to photon-number correlation in the isolated-pair regime is the same regardless of pulse durations. It depends only on the (well-known) fact that for EPP, a photon is always accompanied by one other, whereas for a coherent state one relies on an 'accidental' pairing of photons within the interaction time $T$. [9, 10] The photon-number-correlation part of the quantum enhancement factor Eq.(53) can be written in terms of the mean number of entangled photons from a quantum light source $N_{EPP}$ (necessarily less than one) and the mean number of coherent photons from a laser light source $N_{coh}$ within the same interaction time:

$$QEF_{number} = \frac{N_{EPP}}{N_{coh}^2} \qquad (70)$$

For $N_{EPP} = N_{coh} \ll 1$, this factor can be large, but in this regime the rate of TPA-induced fluorescence counts is extremely small, typically undetectable with current technology.

We found the enhancement due to spectral correlation to be expressed in terms of a convolution of the molecular TPA line profile and the diagonally-projected two-photon state of the EPP and that of a comparator coherent state, as in Eq.(55). There is some ambiguity in concluding that large enhancement exists, as it depends on which EPP state is compared to which coherent state. The largest, and more easily understood, enhancement occurs when comparing a broadband EPP state generated by a long quasi-monochromatic pump laser pulse to a long quasi-monochromatic coherent state. This case is illustrated in **Fig.3**.

In contrast, when EPP are generated by SPDC pumped by a sub-ps laser, minimal spectral entanglement is present typically in the two-photon state, and spectral enhancement cannot be significant. Thus, there is not a way to define a single, unique 'enhanced cross section,' because the absorption probability depends in subtle ways on the nature of the two-photon state.

In experiments carried out in our laboratory (to be published elsewhere), the two-photon absorption of typical molecules with isolated EPP is found to be unobservable, [15] even with sophisticated phase-sensitive (lock-in) photon-counting schemes as described in [41]. An independent experimental study reached similar conclusions. [16]

It is worth commenting that there is some indication that enhancement by spectral correlation in 'classical' light (without entanglement) might be able to mimic that predicted for EPP. [42] It seems likely, however, that such classically correlated light would yield unwanted background counts caused by TPA from 'incoherent' interactions as discussed for intense squeezed light by Dayan *et al*. [17, 12] Such questions are beyond the scope of the present paper.

**Acknowledgements**

This work was supported by grants from the National Science Foundation RAISE-TAQS Program (PHY-1839216 to M.G.R., A.H.M., and B.S. as co-PIs) and by the National Science Foundation Chemistry of Life Processes Program (CHE-1608915 to A.H.M.).

**Data Availability**

The data supporting this study are contained within the article.

**APPENDICES**

**Appendix A** – Solutions of Heisenberg picture equations of motion

The molecular operators $\hat{\sigma}_{ij} = |i\rangle\langle j|$ satisfy the commutation relations $[\hat{\sigma}_{ij}, \hat{\sigma}_{mn}] = \hat{\sigma}_{in}\delta_{jm} - \hat{\sigma}_{mj}\delta_{ni}$ .[24, 25] (These operators are related to Fermion raising and lowering operators, which obey $\hat{b}_i^\dagger \hat{b}_j + \hat{b}_j \hat{b}_i^\dagger = \delta_{ij}$, by $\hat{\sigma}_{ij} = \hat{b}_i^\dagger \hat{b}_j$.) The exact, but implicit, solutions to Eqs.(1) are:

$$\hat{\sigma}_{ij}(t) = \hat{\sigma}_{ij}(t_0)e^{i\omega_{ij}(t-t_0)} + i\sum_m \int_{t_0}^t dt' e^{i\omega_{ij}(t-t')}\left(\mu_{jm}\hat{\sigma}_{im}(t') - \mu_{mi}\hat{\sigma}_{mj}(t')\right)\hat{E}(t') \tag{71}$$

The field and molecule operators commute at equal times, so the ordering of these operators here is arbitrary, but once we start making approximations, the ordering will become fixed.

The initial conditions are $\hat{\sigma}_{ij}(t_0) = |i\rangle\langle j|$, and the zero-th-order (in parameters $\mu_{jm}$) solutions are $\hat{\sigma}_{ij}^{(0)}(t) \simeq \hat{\sigma}_{ij}^{(0)}(t_0)\exp(i\omega_{ij}(t-t_0))$. The first-order solutions are:

$$\hat{\sigma}_{mg}^{(1)}(t_2) = \hat{\sigma}_{mg}(t_0)e^{i\omega_{mg}(t_2-t_0)} - i\mu_{gm}\int_{t_0}^{t_2} dt_1 e^{i\omega_{mg}(t_2-t_1)}\hat{\sigma}_{gg}^{(0)}(t_0)\hat{E}(t_1) \tag{72}$$

Note that the (unequal-time) operator ordering is now fixed. The second-order intermediate-state probabilities are:

$$\hat{\sigma}_{mm}^{(2)}(t_3) = \hat{\sigma}_{mm}^{(2)}(t_0) + i\mu_{mg}\int_{t_0}^{t_3} dt_2 \hat{\sigma}_{mg}^{(1)}(t_2)\hat{E}(t_2) + hc$$

$$= \hat{\sigma}_{mm}^{(2)}(t_0) + i\mu_{mg}\int_{t_0}^{t_3} dt_2 \left\{\hat{\sigma}_{mg}(t_0)e^{i\omega_{mg}(t_2-t_0)} - i\mu_{gm}\int_{t_0}^{t_2} dt_1 e^{i\omega_{mg}(t_2-t_1)}\hat{\sigma}_{gg}^{(0)}(t_0)\hat{E}(t_1)\right\}\hat{E}(t_2) + hc \quad (73)$$

We assume that the initial state is $|g\rangle|\Psi\rangle$ for a pure state $|\Psi\rangle$ of the field, or for a mixed field state a density operator: $\hat{\rho} = |g\rangle\langle g| \otimes \hat{\rho}_{Field}$. The mean probability to excite state $m$ is:

$$\rho_{mm}^{(2)}(t_3) = Tr_{Field}\left(\hat{\rho}_{Field}\langle g|\hat{\sigma}_{mm}^{(2)}(t_3)|g\rangle\right) = p_m + cc \quad (74)$$

where:

$$p_m = \mu_{mg}\mu_{gm}\int_{-\infty}^{t_3} dt_2 \int_{-\infty}^{t_2} dt_1 e^{-(\gamma_{mg}-i\omega_{mg})(t_2-t_1)}\left\langle \hat{E}(t_1)\hat{E}(t_2)\right\rangle \quad (75)$$

where the field correlation function is given by the expectation value:

$$C(t_1,t_2) \equiv \left\langle \hat{E}(t_1)\hat{E}(t_2)\right\rangle = Tr_{Field}\left(\hat{\rho}_{Field}\hat{E}(t_1)\hat{E}(t_2)\right) \quad (76)$$

and we incorporated molecular dephasing rates $\gamma_{mg}$ (from elastic collisions, not population damping) according to Kubo's prescription. [23] Note that when dephasing terms are added to Eqs.(1) it is required to add fluctuation (Langevin) operators as well, [43] but these extra terms do not enter into the expressions for populations, or probabilities, which are of interest here.

Changing variables to $\tau = t_2 - t_1$ gives:

$$p_m = \mu_{mg}\mu_{gm}\int_{-\infty}^{\infty} dt_2 \int_0^{\infty} d\tau e^{-(\gamma_{mg}-i\omega_{mg})\tau}C(t_2-\tau,t_2) \quad (77)$$

To evaluate the correlation function, we write the field operator as a sum of positive- and negative-frequency parts, $\hat{E}(t) = \hat{E}^{(+)}(t) + \hat{E}^{(-)}(t)$, where

$$\hat{E}^{(+)}(t) = \int d\omega L(\omega)\hat{a}(\omega)e^{-i\omega t} \quad (78)$$

with $\hat{E}^{(-)} = \hat{E}^{(+)\dagger}$ and where $L(\omega) = \sqrt{\hbar\omega/2\varepsilon_0 ncA_0}$. The commutator for monochromatic creation and annihilation operators is $[\hat{a}(\omega'),\hat{a}^{\dagger}(\omega)] = 2\pi\delta(\omega'-\omega)$.

Using the rotating-wave approximation and Eqs. (5) and (6), the integral in Eq.(77) yields:

$$p_m = \mu_{mg}\mu_{gm} \int_{-\infty}^{\infty} dt_2 \int_0^{\infty} d\tau\, e^{-(\gamma_{mg}-i\omega_{mg})\tau} \left\langle \hat{E}^{(-)}(t_2-\tau)\hat{E}^{(+)}(t_2) \right\rangle$$

$$= \mu_{mg}\mu_{gm} \int d\omega' L(\omega') \int d\omega L(\omega) \left\langle \hat{a}^\dagger(\omega')\hat{a}(\omega) \right\rangle \frac{2\pi\delta(\omega'-\omega)}{\gamma_{mg}-i\omega_{mg}+i\omega'} \quad (79)$$

$$\simeq |\mu_{gm}|^2 L_0^2 \int d\omega \frac{\left\langle \hat{a}^\dagger(\omega)\hat{a}(\omega) \right\rangle}{\gamma_{mg}-i(\omega_{mg}-\omega)}$$

where $L_0 = \sqrt{\hbar\omega_0/2\varepsilon_0 ncA_0}$ and $\omega_0$ is the (central) carrier frequency of the exciting light.

To calculate two-photon absorption, we first solve for the third-order coherences between final and intermediate states:

$$\hat{\sigma}_{fm}^{(3,coh)}(t_4) = \hat{\sigma}_{fm}^{(0)}(t_0)e^{i\omega_{fm}(t_4-t_0)} + i\mu_{mg} \int_{t_0}^{t_4} dt_3\, e^{i\omega_{fm}(t_4-t_3)} \hat{\sigma}_{fg}^{(2)}(t_3)\hat{E}(t_3)$$

$$= -i\sum_{m'} \mu_{mg}\mu_{m'f}\mu_{gm'} \int_{t_0}^{t_4} dt_3\, e^{i\omega_{fm}(t_4-t_3)} \int_{t_0}^{t_3} dt_2\, e^{i\omega_{fg}(t_3-t_2)} \int_{t_0}^{t_2} dt_1\, e^{i\omega_{m'g}(t_2-t_1)} \hat{\sigma}_{gg}^{(0)}(t_0)\hat{E}(t_1)\hat{E}(t_2)\hat{E}(t_3) \quad (80)$$

where we dropped all initial-value operators that will not contribute to final population expectation values. The population $\rho_{ff}(\infty) = \left\langle \hat{\sigma}_{ff}(\infty) \right\rangle$ in state $|f\rangle$ following the exciting pulse is given by the fourth-order expression $\rho_{ff}^{(4)}(\infty) = P_f + cc$, where:

$$P_f \simeq i\mu_{fm} \sum_m \int_{t_0}^{\infty} dt_4 \left\langle \hat{\sigma}_{fm}^{(3,coh)}(t_4)\hat{E}(t_4) \right\rangle \quad (81)$$

and we used the fact that $\hat{\sigma}_{fm}(t_4)$ and $\hat{E}(t_4)$ commute. This expression evaluates to:

$$P_f = \sum_{m,m'} M_{mm'} p_{mm'} \quad (82)$$

where $M_{mm'} = \mu_{fm}\mu_{mg}\mu_{m'f}\mu_{gm'}$ and, using the change of variables $r = t_4-t_3,\ s = t_3-t_2,\ \tau = t_2-t_1,\ t = t_4$, we find, including dephasing rates $\gamma_{ij}$:

$$p_{mm'} = \int_{-\infty}^{\infty} dt \int_0^{\infty} dr \int_0^{\infty} ds \int_0^{\infty} d\tau\, e^{-(\gamma_{fm}-i\omega_{fm})r} e^{-(\gamma_{fg}-i\omega_{fg})s} e^{-(\gamma_{m'g}-i\omega_{m'g})\tau} C^{(4)} \quad (83)$$

where

$$C^{(4)} = \left\langle \hat{E}(t-r-s-\tau)\hat{E}(t-r-s)\hat{E}(t-r)\hat{E}(t) \right\rangle \quad (84)$$

Applying the rotating-wave approximation:

$$\langle \hat{E}(t_1)\hat{E}(t_2)\hat{E}(t_3)\hat{E}(t_4)\rangle \simeq \langle \hat{E}^{(-)}(t_1)\hat{E}^{(-)}(t_2)\hat{E}^{(+)}(t_3)\hat{E}^{(+)}(t_4)\rangle \tag{85}$$

Then, inserting the field operator Eq.(5) yields

$$C^{(4)} \simeq L_0^4 \int d\omega' \int d\tilde{\omega} \int d\omega \int d\tilde{\omega}' e^{i\omega'(t-r-s-\tau)} e^{i\tilde{\omega}'(t-r-s)} e^{-i\omega(t-r)} e^{-i\tilde{\omega}t} C_\omega^{(4)} \tag{86}$$

where:

$$C_\omega^{(4)} = \langle \hat{a}^\dagger(\omega')\hat{a}^\dagger(\tilde{\omega}')\hat{a}(\omega)\hat{a}(\tilde{\omega})\rangle \tag{87}$$

Integrating over the $t$ variable yields a delta function in the four frequencies, consistent with the fact that when two photons are absorbed over a lengthy time interval, their energies must sum to a particular value, as in the double-sided Feynman diagram in **Fig. 1**. The delta function results in the replacement $\tilde{\omega}' \to \omega' + \omega - \tilde{\omega}$, giving:

$$p_{mm'} = L_0^4 \int d\omega \int d\tilde{\omega} \int d\omega' \frac{\langle \hat{a}^\dagger(\omega')\hat{a}^\dagger(\omega+\tilde{\omega}-\omega')\hat{a}(\omega)\hat{a}(\tilde{\omega})\rangle}{(\gamma_{fm}-i\omega_{fm}+i\tilde{\omega})(\gamma_{fg}-i\omega_{fg}+i\omega+i\tilde{\omega})(\gamma_{m'g}-i\omega_{m'g}+i\omega')} \tag{88}$$

which is equivalent to Eq.(12) considering that the exciting field is far from resonance with any intermediate states.

**Appendix B** – One-Photon Absorption

For reference, we give expressions for one-photon absorption for examples of states of the exciting field: coherent state, single-photon state, and low-flux (isolated) EPP.

    **B.1** Coherent state

For a coherent-state $|\alpha\rangle$ creating population in a state denoted $m$, the two-time correlation function Eq.**(3)** evaluates, under the RWA, using $\langle \hat{a}^\dagger(\omega)\hat{a}(\omega)\rangle = \alpha^*(\omega)\alpha(\omega)$, to:

$$\begin{aligned}C(t_1,t_2) &\simeq \langle \alpha|\hat{E}^{(-)}(t_1)\hat{E}^{(+)}(t_2)|\alpha\rangle \\ &= L_0^2 A^*(t_1) A(t_2) e^{-i\omega_0(t_2-t_1)}\end{aligned} \tag{89}$$

and the absorption probability is:

$$P_m\big|_{coh} = p_{m,coh} + cc$$

$$= |\mu_{gm}|^2 L_0^2 N \int d\omega |\phi(\omega)|^2 \left( \frac{2\gamma_{mg}}{\gamma_{mg}^2 + (\omega_{mg} - \omega)^2} \right) \tag{90}$$

where we used $\alpha(\omega) = \alpha_0 \phi(\omega)$. This is simply the overlap of the (normalized) exciting spectrum $|\phi(\omega)|^2$ with the absorption profile. Note that the unit of $|\mu_{gm}|^2 L_0^2$ is s$^{-1}$.

If the exciting field is a long narrow-band pulse of coherent-state light, which could be a 'rectangular' pulse chopped from a continuous-wave (CW) beam, we model it by assuming the exciting spectral density $|\phi(\omega)|^2$ is peaked at $\omega_0$ and much narrower than the absorption profile; then:

$$P_m\big|_{coh} \simeq |\mu_{gm}|^2 L_0^2 N \frac{2\gamma_{mg}}{\gamma_{mg}^2 + (\omega_{mg} - \omega)^2} \tag{91}$$

### B.2 Single-photon state

For a single-photon state $|\varphi\rangle$, Eq.(18), creating probability in state $m$, the correlation function is:

$$C_{coh}(t_1, t_2) = L_0^2 \varphi^*(t_1) \varphi(t_2) e^{-i\omega_0(t_2 - t_1)} \tag{92}$$

where the time-domain wave packet is:

$$\varphi(t) \equiv \int d\omega \tilde{\varphi}(\omega) e^{-i(\omega - \omega_0)t} \tag{93}$$

By the same calculation as for the coherent state, we get:

$$P_m\big|_{1\,photon} = |\mu_{gm}|^2 L_0^2 \int d\omega |\tilde{\varphi}(\omega)|^2 \left( \frac{2\gamma_{mg}}{\gamma_{mg}^2 + (\omega_{mg} - \omega)^2} \right) \tag{94}$$

which is identical to Eq.(90) when $N$ is replaced by one.

Therefore, when exciting a single molecule in the linear-response regime, a single-photon pulse has the same effect as a coherent-state pulse with mean photon number equal to one. In this scenario there is nothing especially 'quantum' about single-photon absorption. The story is different when exciting two or more molecules—the single-photon pulse can create entanglement of excitation in this case, whereas a coherent-state pulse does not. Such entanglement is the basis for atomic-ensemble quantum memories, which store states of light in an extended 'phased-array' of atoms.

**Appendix C** – TPA by a coherent pulse in the impulsive limit

In an extreme limit, if the coherent state is a pulse much shorter than the molecular dephasing time—the impulsive limit—then Eq.(23) becomes:

$$P_f = \Sigma^{(2)} L_0^4 \left| \int A^2(t) dt \right|^2 \qquad (95)$$

where we used $K(0) = \int A^2(t) dt$.

If A(t) is real and positive, then $K(0) = 1$, reproducing the result in Eq.26 for the Gaussian coherent state in the impulsive limit $\sigma \gg \gamma_{fg}$. In this limit the probability does not depend on $\gamma_{fg}$ or $\sigma$ because for the nonlinear TPA process the effect of spreading the spectrum over a broader range is compensated by the increasing intensity in the time domain.

Note that $A^2(t)$ is proportional to the two-photon Rabi frequency [44] and can be complex. This means that the TPA probability can go to zero in the impulsive limit if the pulse constitutes a two-photon zero-$\pi$ pulse, defined by $K(0) = 0$.

**Appendix D** – TPA with long coherent-state pulse

To derive Eq.(27) for a long quasi-monochromatic pump pulse, we go back to the earlier form Eq.(14), which for the coherent state is well approximated by:

$$p_f = \frac{\Sigma^{(2)} L_0^4}{\gamma_{fg} - i\omega_{fg} + i2\omega_0} Int \qquad (96)$$

where

$$\begin{aligned} Int &= \int d\omega \int d\tilde{\omega} \int d\omega' \alpha^*(\omega') \alpha^*(\omega + \tilde{\omega} - \omega') \alpha(\omega) \alpha(\tilde{\omega}) \\ &= \int |A(t)|^4 dt \end{aligned} \qquad (97)$$

where the final step was obtained by substituting the Fourier transform Eq.(16).

**Appendix E** – TPA by Type-II EPP

For Type-II SPDC, where the signal (*S*) and idler (*I*) modes are distinct, the annihilation operator in Eq.(5) is replaced by $\hat{a}(\omega) = \hat{a}_S(\omega) + \hat{a}_I(\omega)$. The state is [45]:

$$|\Psi\rangle = \sqrt{1-\varepsilon^2}|vac\rangle + \varepsilon \int d\omega \int d\tilde{\omega}\, \psi(\omega,\tilde{\omega})\hat{a}_S^\dagger(\omega)\hat{a}_I^\dagger(\tilde{\omega})|vac\rangle + \ldots \qquad (98)$$

The correlation function is expressed again as $\tilde{C}_\omega = \varphi^*(\omega',\tilde{\omega}')\varphi(\omega,\tilde{\omega})$, where:

$$\begin{aligned}
\varphi(\omega,\tilde{\omega}) &= \langle vac|\hat{a}(\omega)\hat{a}(\tilde{\omega})|\Psi\rangle \\
&= \varepsilon \langle vac|(\hat{a}_S(\omega)+\hat{a}_I(\omega))(\hat{a}_S(\tilde{\omega})+\hat{a}_I(\tilde{\omega}))\int d\omega_1 \int d\omega_2\, \psi(\omega_1,\omega_2)\hat{a}_S^\dagger(\omega_1)\hat{a}_I^\dagger(\omega_2)|vac\rangle \\
&= \varepsilon\psi(\omega,\tilde{\omega}) + \varepsilon\psi(\tilde{\omega},\omega)
\end{aligned} \qquad (99)$$

In order to use for Type-II SPDC the results that were derived for Type-I, we assume that all the light illuminating the molecule has a common polarization, so the coupling coefficients can still be written $\mu_{ij} = \vec{d}_{ij} \cdot \vec{\mathcal{E}}/\hbar$. This can be aranged in experiments by separating the signal and idler beams, rotating one of their polarizations and combining the beams at a small angle in the sample.

Thus, Eq.(32) still holds, with the replacement of $\psi(\omega,\tilde{\omega})$ with $\Psi(\omega,\tilde{\omega}) = \{\psi(\omega,\tilde{\omega}) + \psi(\tilde{\omega},\omega)\}/2$, which has boson symmetry, and acts like a two-photon detection amplitude (or two-photon wave function [46]), and TPA acts like a two-photon detector.

**Appendix F** – TPA by a short EPP pulse in the impulsive limit

If the EPP state is a pulse much shorter than the molecular dephasing times—the impulsive limit—then Eq.(32) becomes:

$$\begin{aligned}
P_f &= \Sigma^{(2)} 4\varepsilon^2 L_0^4 \left| \int d\omega\, \psi(\omega, 2\omega_0 - \omega) \right|^2 \\
&= \Sigma^{(2)} 4\varepsilon^2 L_0^4 \left| \int \tilde{\psi}(t,t)\, dt \right|^2
\end{aligned} \qquad (100)$$

where:

$$\tilde{\psi}(t,\tilde{t}) = \int d\tilde{\omega}\, e^{-i(\tilde{\omega}-\omega_0)\tilde{t}} \int d\omega\, e^{-i(\omega-\omega_0)t} \psi(\omega,\tilde{\omega}) \qquad (101)$$

Equation (100) is the EPP generalization of the impulsive coherent-state result Eq.(95). As such, we can interpret $\tilde{\psi}(t,t)$ as being proportional to a 'quantum two-photon Rabi frequency,' implying that the TPA probability can go to zero in the impulsive limit if the EPP pulse constitutes a two-photon zero-$\pi$ pulse.

**Appendix G** – TPA cross section

Define the optical flux density ($photons\ m^{-2}\ s^{-1}$) in a coherent-state beam as $I = (2n\varepsilon_0 c / \hbar\omega)\,|A(t)|^2$, where the field amplitude is defined in Eq.(15). The relation to flux ($photons\ s^{-1}$) is $I = F_{coh}/A_0$, where $A_0$ is the effective beam area. The rate of TPA is defined to be $R = \sigma^{(2)} I^2$, where $\sigma^{(2)}$ is the two-photon cross section. On two-photon resonance, the population in state $f$ after interaction for time $T$ with a CW beam is, from Eq.(27):

$$P_f\Big|_{coh} = 2\Sigma^{(2)} \left(\frac{\hbar\omega_0}{2\varepsilon_0 n c A_0}\right)^2 \frac{1}{\gamma_{fg}} F_{coh}^{\ 2} T \qquad (102)$$

Combining these formulas using $P_f = RT$ yields:

$$\begin{aligned}
\sigma^{(2)} &= 2\Sigma^{(2)} L_0^{\ 4} \frac{A_0^{\ 2}}{\gamma_{fg}} \\
&= \left(\frac{\omega_0}{\hbar\varepsilon_0 n c}\right)^2 \frac{1}{2\gamma_{fg}} \sum_{m,m'} \frac{d_{fm} d_{mg} d_{m'f} d_{gm'}}{(-\omega_{fm} + \omega_0)(\omega_{m'g} - \omega_0)}
\end{aligned} \qquad (103)$$

where $d_{ij} = \vec{d}_{ij} \cdot \vec{\varepsilon}$. This result is within a factor 4 of the conventional one derived using the concept of density of states. Note that it depends on no beam parameters other than its frequency.